# Universal Scaling in Non-equilibrium Transport Through a Single-Channel Kondo Dot

M. Grobis,[1] I. G. Rau,[2] R. M. Potok,[1,3,*] H. Shtrikman,[4] and D. Goldhaber-Gordon[1]

[1] Department of Physics, Stanford University, Stanford, CA 94305

[2] Department of Applied Physics, Stanford University, Stanford, CA 94305

[3] Department of Physics, Harvard University, Cambridge, MA 02138

[4] Department of Condensed Matter Physics, Weizmann Institute of Science, Rehovot 96100, Israel

Scaling laws and universality play an important role in our understanding of critical phenomena and the Kondo effect. Here we present measurements of non-equilibrium transport through a single-channel Kondo quantum dot at low temperature and bias. We find that the low-energy Kondo conductance is consistent with universality between temperature and bias and characterized by a quadratic scaling exponent, as expected for the spin-½ Kondo effect. We show that the non-equilibrium Kondo transport measurements are well-described by a universal scaling function with two scaling parameters.

Seemingly unrelated physical phenomena can sometimes display strikingly similar behavior. The evolutions of thermodynamic parameters of certain liquid-gas and paramagnetic-ferromagnetic phase transitions, for example, are characterized by identical sets of critical exponents even though the underlying forces (van der Waals and magnetic exchange, respectively) are vastly different [1]. A similar type of universality can also be exhibited in the response of a physical system to different perturbations such as temperature, applied bias, or magnetic field. Here, even though each perturbation affects the system in a qualitatively different manner, the characteristic exponent of the lowest order response is identical for several of the perturbations. While the coefficients of the lowest order response generally depend on system-specific energy scales, these dependences can usually be eliminated by scaling each perturbation relative to a characteristic energy, allowing the whole class of systems to be neatly described by a single universal scaling function.

The universal scaling laws governing perturbations that drive a system out-of-equilibrium have been especially difficult to derive theoretically with high accuracy. One of the best-known phenomena that has eluded a precise theoretical description is the non-equilibrium Kondo effect [2-9]. The Kondo effect is a many-body state that arises from the interaction of a local magnetic moment with a reservoir of conduction electrons [10-12]. The effect has been observed in a variety of systems including mesoscopic quantum dots, where the spin of an unpaired electron on the dot acts as the magnetic impurity [13, 14]. For quantum dots, the non-equilibrium Kondo effect can occur whenever two reservoirs (i.e. two leads) participate in forming the Kondo state. Applying a bias voltage between the two leads drives the Kondo state out-of-equilibrium as neither reservoir is in equilibrium with the quantum dot. While previous measurements have shown that transport at zero bias does indeed follow a universal scaling curve in temperature, the low energy power law temperature and non-equilibrium regimes have not been closely examined [15]. All theoretical considerations predict that universal scaling behavior in temperature $T$ and applied bias $V$ is maintained in the non-equilibrium Kondo regime [3, 4, 11], but they disagree about the coefficients values of the universal scaling function and how many system-specific parameters determine the scaling relations [6]. Experimental work has not elucidated the theoretical situation, as existing measurements have focused on higher energies ($eV \sim kT_K$) from which reliable low-energy scaling behavior cannot be extracted [13-17].

In this Letter we examine low temperature and low bias transport through a spin-½ Kondo dot in the non-equilibrium regime. We find that the non-equilibrium conductance in the Kondo plateau is consistent with a quadratic power-law at low energies [18], as previously predicted for single-channel Kondo transport but not examined experimentally. Furthermore we show that low energy conductance is well-described by a universal scaling function in temperature and bias with two scaling parameters: the Kondo temperature $T_K$ and the zero temperature conductance $G_0$. Our extracted universal scaling function coefficients are in general agreement with calculations within the Anderson-model framework. An explanation of the observed deviations will require further study.

Our measurements are performed using a quantum dot formed from a low density ($n_e = 2 \times 10^{11}$ e$^-$/cm$^2$), high mobility ($\mu = 2\times10^6$ cm$^2$/V s) two dimensional electron gas (2DEG) located 68 nm below the surface of a GaAs/AlGaAs heterostructure. Gate electrodes (Fig. 1(a)) define the dot and finely tune its coupling to extended regions of 2DEG, which serve as leads. One gate in particular, labeled $V_G$ in Fig. 1(a), tunes the quantum dot to odd occupancy and controls the energy $\varepsilon_0$ of the singly occupied level while minimally perturbing the dot-lead coupling $\Gamma$. The dot's area is approximately 0.04 μm$^2$, leading to a bare charging energy $U \sim 1$ meV and mean level spacing Δ ~ 100 μeV [19]. We measure differential conductance $G = dI/dV$ using standard lock-in techniques, with an rms modulation bias of 1 μV at 337 Hz. The dot is cooled in the mixing chamber of an Oxford Instruments Kelvinox TLM dilution refrigerator with a base electron temperature of 13 mK [20].

At base temperature, the Kondo effect produces a characteristic enhancement of conductance through the quantum dot at odd occupancy ($-\varepsilon_0$, $\varepsilon_0 + U > \Gamma$), as seen in Fig. 1. The couplings of the dot to its two leads are tuned to maximize $T_K$ while keeping the two couplings nearly equal. The saturation of conductance at a value near 1.75 $e^2/h$ throughout the middle of the Kondo plateau (-209 mV < $V_G$ < -199.5 mV in Fig. 1(d)) confirms that $T_{base} \ll T_K$ and indicates that the coupling asymmetry is around 2:1 [2]. Conductance as a function of source-drain bias in the Kondo plateau shows a narrow peak centered at zero bias, known as the Kondo peak (Fig. 1(c)).

As the temperature is increased from 13 mK to 205 mK the overall Kondo conductance decreases (Fig.1 (c,d)). Previous measurements have found that the temperature evolution of linear conductance $G(T,V=0)$ is well-described by an empirical Kondo (EK) form derived from a fit to numerical renormalization group (NRG) conductance calculations [15],

$$G_{EK}(T) = G_0/(1 + (T/T_K')^2)^s, \qquad (1)$$

Here, $s = 0.21$ and $T_K' = T_K/(2^{1/s} - 1)^{1/2}$, which defines $T_K$ as the temperature at which the Kondo conductance is half of its extrapolated zero temperature value: $G_{EK}(T_K) = G_0/2$. The values of $G_0$ and $T_K$ we extract are shown in Fig. 2(a). The Kondo conductance traces in our measurements follow the EK form very well at low temperatures ($T < T_K/4$), but deviate from the form at higher temperatures, as seen in Fig. 2(b). Though the origin of the deviation at higher temperatures is not completely understood, it most likely reflects the emergence of additional transport processes at higher temperatures. In extracting the Kondo fitting parameters we limit the temperature range for our fits to $T < 35$ mK $\approx T_K/4$ in the middle of the plateau.

To determine whether bias and temperature obey a scaling relationship at low temperatures we fit the low energy conductance to the form

$$G(T,V) \approx G_0 - \tilde{c}_T(kT)^{P_T} - \tilde{c}_V(eV)^{P_V} . \qquad (2)$$

Here $P_T$ and $P_V$ are exponents that characterize the temperature and bias dependence, respectively, and $\tilde{c}_T$ and $\tilde{c}_V$ are expansion coefficients. Unlike the EK form (Eq. (1)), Eq. (2) does not assume quadratic behavior at low temperature. We first extract $P_V$ by fitting $G(T,0) - G(T,V)$ as a power-law in voltage for $|V| < 7$ µV at each temperature point below 20 mK. We find that $P_V$ is nearly constant across the Kondo plateau with an average value of $1.9 \pm 0.15$ (Fig. 3(a)) [21], in good agreement with the predicted single-channel Kondo exponent of 2. Extracting $P_T$ is more difficult since at each gate voltage point only a few temperature points unambiguously reside in the power-law regime ($T/T_K \lesssim 0.1$), as seen in Fig. 2(b). Fitting over this temperature range yields a mean value of $P_T = 2.0 \pm 0.6$ across the Kondo plateau [21]. These fits are consistent with temperature and bias sharing a characteristic exponent of 2, as theoretically expected for the single-channel Kondo effect, and we assume this universality for the remainder of our analysis.

Having determined the characteristic scaling exponent, we now examine to what extent the low energy non-equilibrium conductance $G(T,V)/G_0$ is described by a universal scaling function, $F(T/T_K, eV/kT_K)$. We assume as a starting point that $G(T,0)$ follows the universal curve given by Eq. (1) and examine the evolution of the differential conductance with temperature at finite bias. For each point in the Kondo plateau we fit the Kondo peak using the low bias expansion of a form that Refs. [22] found applicable over a wide temperature range:

$$G(T,V) = G_{EK}(T,0)\left(1 - \frac{c_T \alpha}{1 + c_T\left(\frac{\gamma}{\alpha} - 1\right)\left(\frac{T}{T_K}\right)^2}\left(\frac{eV}{kT_K}\right)^2\right). \tag{3}$$

The coefficient $c_T \approx 5.49$ is fixed by the definition of $T_K$ via Eq. (1): $G_{EK}(T,0) \approx G_0\left(1 - c_T\left(\frac{T}{T_K}\right)^2\right)$. The coefficients $\alpha$ and $\gamma$ characterize the zero-temperature curvature and temperature broadening of the Kondo peak, respectively, and are defined appropriately to be independent of how $T_K$ is defined. The form of Eq. (3) is chosen so that at low temperature Eq. (3) reduces to a universal scaling function expansion suggested by Schiller *et al.* [4]:

$$\frac{G(T,0) - G(T,V)}{c_T G_0} = F\left(\frac{T}{T_K}, \frac{eV}{kT_K}\right) \approx \alpha\left(\frac{eV}{kT_K}\right)^2 - c_T\gamma\left(\frac{T}{T_K}\right)^2\left(\frac{eV}{kT_K}\right)^2 \tag{4}$$

Figures 3(b) and 3(c) show the extracted values of α and γ across the Kondo plateau, fitted over the regime $T/T_K < 0.2$ and $eV/kT_K < 0.4$. The coefficients are nearly constant across the Kondo plateau and have average values $\alpha = 0.10 \pm 0.015$ and $\gamma = 0.5 \pm 0.1$ in the middle of the Kondo plateau (-209 mV ≤ $V_G$ ≤ 199.5 mV) [21, 23]. Both $\alpha$ and $\gamma$ increase slightly on the right side of the Kondo plateau ($V_G$ > -199.5 mV). This may reflect the expected break-down of the universal scaling relation in the regime where charge fluctuations modify Kondo processes (mixed valence regime).

Overall, we can conclude that low energy transport through a Kondo dot in the Kondo regime is well-described by the universal scaling function given by Eqs. (3) and (4). The degree of agreement can be visualized by plotting the scaled conductance $(1 - G(T,V)/G(T,0))/\tilde{\alpha}_V$ versus $(eV/kT_K)^2$, where $\tilde{\alpha}_V = c_T\alpha / \left(1 + c_T\left(\frac{\gamma}{\alpha} - 1\right)\left(\frac{T}{T_K}\right)^2\right)$, using the average values of $\alpha$ and $\gamma$ across the Kondo plateau. As Figure 4 shows, the conductance data across the Kondo plateau for $T/T_K < 0.6$ collapse nicely onto a single universal curve for bias values up to $(eV/kT_K)^2 \sim 0.5$, above which deviations due to higher-order terms or non-Kondo processes become apparent.

Previous experimental work on spin-½ Kondo dots mainly examined the full-width at half-maximum (FWHM) of the Kondo peak at fixed values of $T_K$, rather than the whole Kondo plateau, and did not investigate the low bias power-law regime [14, 16, 22, 24]. For comparison purposes we extract rough values of $\alpha$ and $\gamma$ from a few existing measurements by approximating the Kondo peak as a Lorentzian. van der Wiel *et al.*'s measurement of Kondo transport in the non-equilibrium regime gives $\alpha \approx 0.25$ and $\gamma \approx 0.5$ in the middle of the Kondo plateau [16]. From transport measurements through magnetic impurities coupled to highly asymmetric leads [25], which probe the equilibrium rather than non-equilibrium spectral function, we extract $\alpha \approx 0.05$ and $\gamma \approx 0.1$ in Refs. [22] and $\alpha \approx 0.15$ and $\gamma \approx 0.5$ in Ref. [26]. Though these values are roughly comparable to our measured values, the wide variation they exhibit emphasizes the importance of focusing on low energies in order to extract meaningful coefficients.

We now compare our carefully extracted values for $\alpha$ and $\gamma$ to existing theory for non-equilibrium transport through a spin-½ Kondo quantum dot. Existing calculations are based on either the Anderson [6, 9, 11] or Kondo models [4, 5, 12], and focus mainly on determining $\alpha$. Quantum dot measurements have been generally well-described by the Anderson model. Calculations using this framework predict a value of 0.15 for $\alpha$ in both the strong coupling non-equilibrium [6] and equilibrium limits [12]. Other theoretical treatments show a greater level of disagreement with our results [5, 8, 9]. A comparison to our measured value of ~0.1 indicates that the observed Kondo conductance decreases with bias slightly more slowly than predicted by the Anderson model calculations. The origin of this discrepancy is currently not understood. One possibility is that the non-equilibrium calculations overestimate how quickly Kondo-processes diminish with increasing bias. A second possibility is that additional non-Kondo transport processes, such as inelastic co-tunneling [27], contribute to a concurrent bias-dependent conductance in our measurement. The contribution, if any, of additional transport processes could be resolved in future experiments by extracting scaling parameters at different dot ($U$, $\Delta$) and coupling ($\Gamma$) parameters for identical $T_K$ and $G_0$ values.

We have performed detailed measurements of the Kondo conductance through a quantum dot as a function of temperature and bias. We find that our non-equilibrium Kondo conductance measurements are consistent with universality in temperature and bias characterized by a quadratic scaling exponent. The conductance is well-described by a universal non-equilibrium Kondo scaling function whose coefficients are constant along the Kondo plateau. The extracted coefficient values show general agreement with Anderson-model calculations and provide a well-defined reference point for further experimental and theoretical work.

We thank N. Andrei, J. Bauer, T. Costi, A. Hewson, J.E. Moore, A. Oguri, Y. Oreg, A. Schiller, and J. von Delft for discussions. This work was supported by NSF CAREER Award

DMR-0349354 and US-Israel BSF Award #2004278. D.G.-G. acknowledges Fellowships from the Sloan and Packard Foundations, and a Research Corporation Research Innovation Award.

**Figure Captions**

**FIG. 1 (a)** The SEM image shows the quantum dot device with an overlaid measurement schematic. The topmost lead (marked "NC") is pinched off from the quantum dot and does not contribute to transport. **(b)** dI/dV measured as a function of $V_G$ and source-drain bias at $T = 13$ mK. **(c)** Temperature dependence of the Kondo peak in conductance for $T = 13$-$205$ mK at $V_G = -203$ mV. **(d)** Temperature dependence of the Kondo plateau for $T = 13$-$205$ mK at $V = 0$ µV.

**FIG. 2 (a)** Values of $G_0$ and $T_K$ across the Kondo plateau, extracted using the empirical Kondo (EK) form $G_{EK}(0,T) = G_0/(1-(T/T_K')^2)^s$ with $T_K' = T_K/(2^{1/s} - 1)^{1/2}$. The fit was performed using data points for temperatures between 13-35 mK. **(b)** A plot of the scaled conductance $1 - G(T,0)/G_0$ versus $T/T_K$ for all measured temperatures and gate voltage points across the Kondo plateau. The solid line shows the empirical Kondo form.

**FIG. 3 (a)** Values of the Kondo scaling exponent for bias ($P_V$) across the Kondo plateau. The horizontal dashed line shows the theoretically predicted single-channel Kondo exponent, $P = 2$. Values of **(b)** $\alpha$ and **(c)** $\gamma$ extracted across the Kondo plateau. The fitting methods are described in the text and in Ref. [21].

**FIG. 4 (a)** Conductance as a function of bias for $0 < T/T_K < 0.6$ for six points across the Kondo plateau (-209 mV < $V_G$ < -199.5 mV). **(b)** Scaled conductance $(1 - G(T,V)/G(T,0))/\tilde{\alpha}_V$ versus $(eV/kT_K)^2$, where $\tilde{\alpha}_V = c_T\alpha / \left(1 + c_T\left(\frac{\gamma}{\alpha} - 1\right)\left(\frac{T}{T_K}\right)^2\right)$. We use $\alpha = 0.10$ and $\gamma = 0.5$, which are the average values of the extracted scaling coefficients in the middle of the Kondo plateau. The solid line shows the associated universal curve described by Eq. (3).

Figure 1

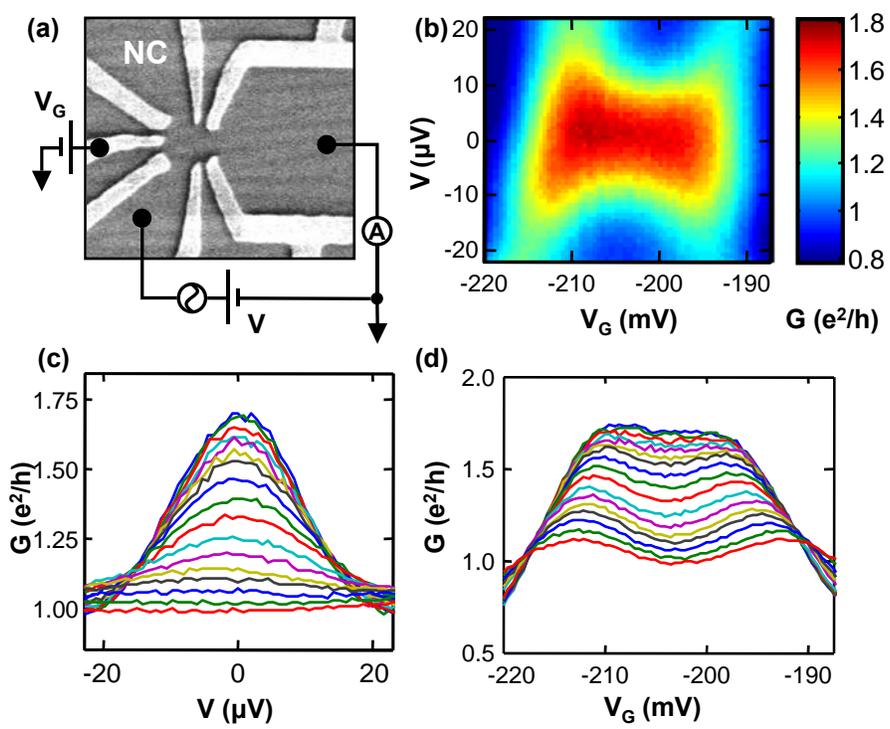

Figure 2

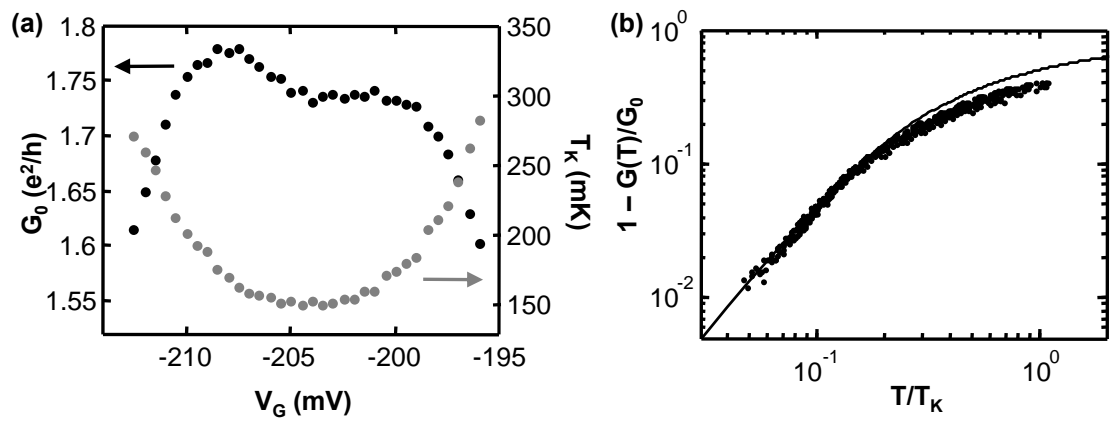

Figure 3

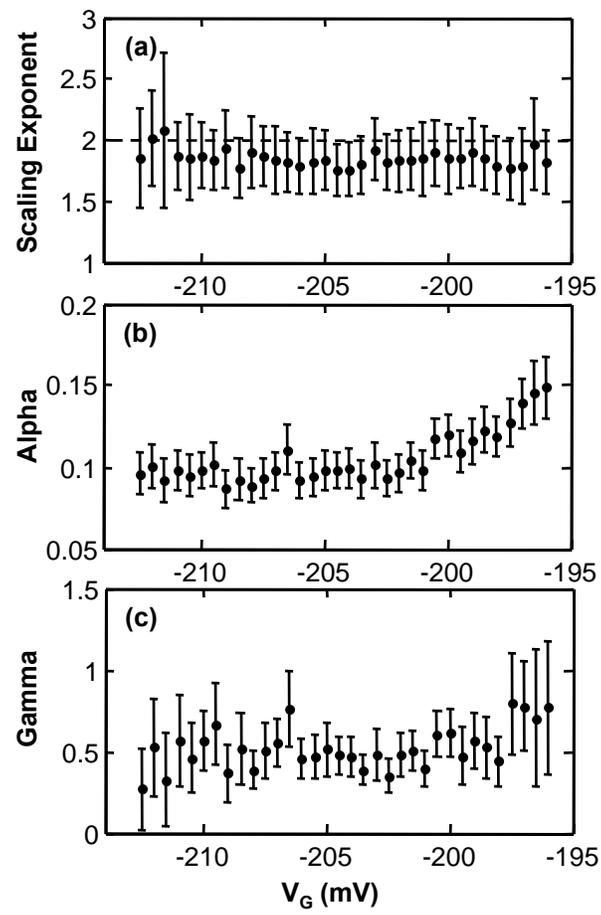

Figure 4

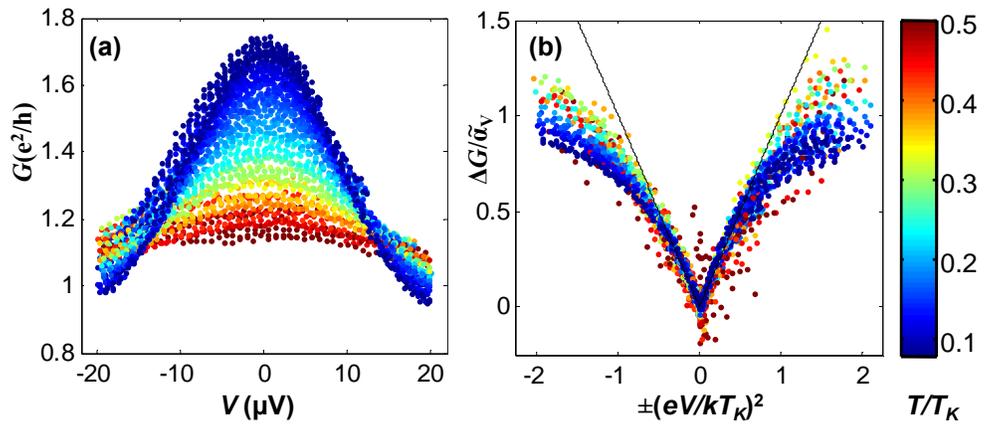

# Universal Scaling in Non-equilibrium Transport Through a Single-Channel Kondo Dot
# (EPAPS Supplementary Info)

M. Grobis,[1] I. G. Rau,[2] R. M. Potok,[1,3,*] H. Shtrikman,[4] and D. Goldhaber-Gordon[1]

[1] Department of Physics, Stanford University, Stanford, CA 94305

[2] Department of Applied Physics, Stanford University, Stanford, CA 94305

[3] Department of Physics, Harvard University, Cambridge, MA 02138

[4] Department of Condensed Matter Physics, Weizmann Institute of Science, Rehovot 96100, Israel

## 1    Introduction

In the main text of our letter we examine the low temperature and bias scaling behavior of Kondo dot transport. Our measurements in the power-law energy regime are consistent with a universal scaling picture in which temperature and bias share a quadratic exponent. Furthermore, we find that the conductance is well-described by a two parameter universal scaling function. In this section we provide additional information about our fitting methods and error bars. We also discuss the effects of fitting $G_0$ and $T_K$ using a temperature-independent offset, a technique commonly used in the experimental literature.[1-3]

## 2    Fitting Range

In order to achieve meaningful fits, the range of bias and temperature points has to be carefully limited in order to avoid including points which are outside the range of validity for each respective fitting function. This section discusses the logic behind the temperature and bias ranges we use for the fits of $G_0$ and $T_K$, the scaling exponent, and the scaling coefficients.

## 2.1   $G_0$ and $T_K$ fits

We use the empirical Kondo form (Eq. 1 of the main text) to determine the values of $G_0$ and $T_K$ across the Kondo plateau. In order to improve the accuracy of the fit, we first fit each Kondo peak using a parabola and use the maximum of the parabolic fit as the zero-bias conductance. This eliminates artifacts stemming from the slow drift (<1 µV per hour) of the input bias of our current amplifier. The empirical Kondo (EK) form accurately fits our extrapolated zero bias conductance values up to $T/T_K \approx 0.25$ (40-60 mK). Fitting above these temperatures creates a significantly worse fit, marked by an increase in the deviations between the data and fit function over the fitting range.  For fits up to $T/T_K \approx 0.25$, the deviations are consistent with experimental noise.

We note that the fit at higher temperatures is improved when a temperature-independent offset conductance is added to the EK form as a fitting parameter. Similar offsets are found in much of the experimental Kondo dot literature. Though the origin of this offset is not completely understood, it most likely is not truly temperature-independent but rather reflects the emergence of additional transport processes at higher temperatures. The quantitative effect of including the offset ($\approx 0.4 \pm 0.1$ $e^2/h$ across the plateau) is a decrease in the best fit $G_0$ and $T_K$ by 20-30%, each, for all points in the Kondo plateau. However, the quality of the fit at low temperatures (< 40mK) is worsened by using an offset: the deviations between the fit and data in the offset fit are twice as large as in the offset-less fit.  Since our work focuses on precisely this low energy regime, we extract $G_0$ and $T_K$ using the offset-less fit, as described above.

## 2.2   Scaling Exponent Fits

Picking the appropriate fit range for extracting the scaling exponents is crucial, since the Kondo conductance seamlessly transitions away from power-law behavior with increasing energy.  Fitting too far out in energy introduces artifacts from higher order terms. We determined the extent of the power-law regime by examining the behavior of $G_0 - G(0,T)$ vs. $T/T_K$ and $G(0,T) - G(V,T)$ vs. $eV/kT_K$ on log-log plots, as shown in Figure S1(a) and S1(b).

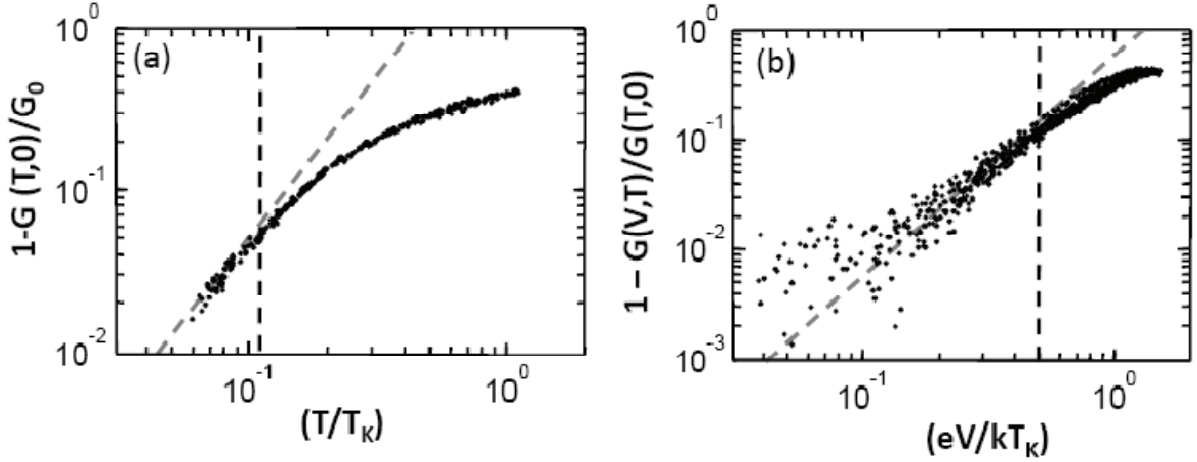

**Figure S1 (a)** Scaled conductance $1 - G(T,0)/G_0$ versus $T/T_K$ for all measured temperatures and for all gate voltage points across the whole Kondo plateau ($-213 < V_G < -196$). The gray dashed line shows $c_T(T/T_K)^2$, which is the low temperature limit of $1 - G(T,0)/G_0$ for the empirical Kondo form ($c_T \approx 5.49$). The vertical line shows the value of $T/T_K$ at which the deviations of the conductance away from the quadratic fit are larger than the noise. **(b)** Scaled conductance $1 - G(T,V)/G(T,0)$ versus $eV/kT_K$ for $T = 13$ mK and $V = \pm 20$ μV for all points across the Kondo plateau ($-209 < V_G < -199.5$). The gray dashed line shows $c_T(\alpha - \gamma c_T(T/T_K))(eV/kT_K)^2 \approx 0.12$, which is the average bias coefficient in the Kondo plateau at base temperature ($T = 13$ mK).

Since we anticipate quadratic exponents, as predicted for single channel Kondo, we limit our fits to regions where the conductance traces follow quadratic behavior (dotted line) to within the scatter of our measured points ($T/T_K \approx 0.11$ and $eV/kT_K \approx 0.5$, as noted by the vertical dashed lines). Though our choice of fit range is guided by assuming single-channel Kondo behavior, the exponent fits themselves make no assumption about the form of the conductance, including the values of $G_0$ and $T_K$. For the temperature exponent fits we only consider gate voltage points for which at least five temperature points fall below $0.11*T_K$. The bias range is further truncated to avoid fitting over regions where the bare Coulomb blockade peak (single-particle resonance) contributes to conductance at finite bias. This occurs near the edges of the Kondo plateau ($V_G \approx -209$ and $V_G \approx -200$). A schematic of the actual bias fit range is shown in Fig. S2.

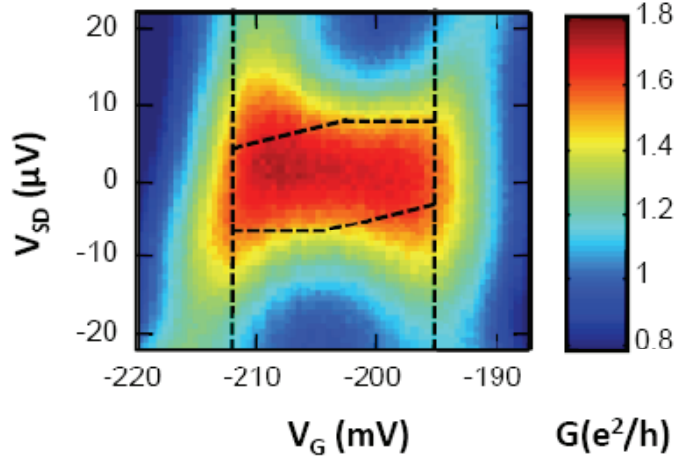

**Figure S2.** The dashed lines show the limits of bias range used for fitting the scaling exponent and scaling coefficient. The vertical dashed lines mark edges of the Kondo plateau points we used in our exponent fits.

### 2.3 Scaling Coefficient Fits

In extracting the scaling coefficients, we assume that the Kondo conductance is a quadratic in voltage of the form $G(T,V) = G(T,0) * \left(1 - \left(\frac{V}{W(T)}\right)^2\right)$, as described by Eq. 3 of the main text. We use the same bias range for the coefficient fits. As Fig. S3 shows, $W(T)^2$ follows a quadratic of the form $aT_K^2 + bT^2$ for a wide range of temperatures. We limit the temperature range for extracting the coefficients to points at which the zero bias conductance follows the EK form ($T/T_K \lesssim 0.25$).

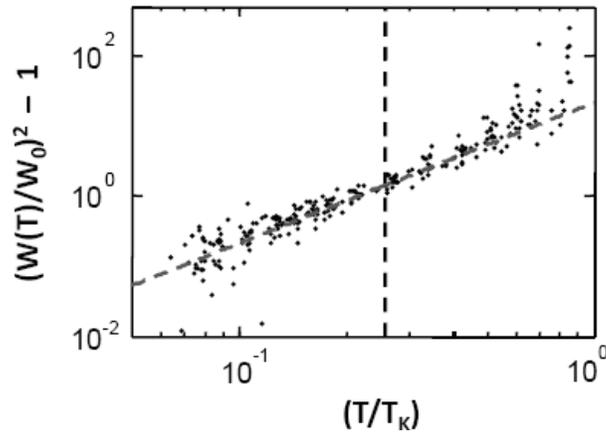

**Figure S3.** Log-log plot of the Kondo peak width parameter $(W(T)/W_0)^2 - 1$ versus scaled temperature $T/T_K$ across the Kondo plateau. For the zero temperature value we use the plateau average $W_0^2 = (1/c_T\alpha) \approx 1.85$. The vertical dashed line marks the limit of the temperature range used to extract the scaling coefficients. The gray dashed represents $c_T\left(\frac{\gamma}{\alpha} - 1\right)\left(\frac{T}{T_K}\right)^2 \approx 21\left(\frac{T}{T_K}\right)^2$, the average value for the scaled temperature broadening in the Kondo plateau.

## 3 Meaning of Error Bars

The error bars in our manuscript represent 68% confidence intervals for the extracted parameters. The error bars stem from statistical errors and two systematic errors: uncertainty in temperature calibration and choice of fitting range in bias and temperature. Statistical errors reflect random measurement noise and contribute mainly to error bars at each individual gate voltage point. The error bars for the average exponent and coefficient values across the plateau arise mainly from the systematic errors, which do not average out across the plateau. In determining the systematic errors we assume an accuracy of ± 0.7 mK for our quoted temperatures, which we calibrated by measuring the widths of temperature broadened Coulomb blockade peaks. We assume the temperature error at each temperature point is random. Though the bias and temperature fit ranges are chosen carefully, as discussed earlier, the extracted exponents and coefficients do depend slightly on the fit range. We take this dependence into consideration by examining the variation caused by changing the fit ranges by one data point in each direction. Hence, the error bars reflect the effects of varying the bias fit range by ± 13% and the temperature fit range by ± 20%. We determined the scaling coefficient error bars by treating the Kondo peak as a quadratic, as discussed above. The effects of using other fitting methods, which are discussed in Sec. 4, are not included in our error bar calculation.

## 4 Effects of Other Peak Fitting Techniques on Extracted Coefficients

### 4.1 Kondo Peak Fits

In extracting the temperature-dependent width of the Kondo peak we use a simple quadratic line shape for the tip of the peak. We choose a quadratic form for its simplicity, given our goal of studying the quadratic energy regime, even though several other fitting functions that reduce to a quadratic at low bias match our data over a wider bias range. For each choice of fitting function, the extracted coefficients always remain nearly constant across the Kondo plateau, like in Fig. 3 of the main text, providing further support that these coefficients are indeed universal.

We note that two forms that gave the best fits for wider ranges of bias were the

Lorentzian $G(T,V) = G(T,0)/\left(1 + \left(\frac{V}{W(T)}\right)^2\right)$ and the modified quadratic $G(T,V) = G(T,0) *$
$(1 - V^2/(W(T)^2 + Q(T)^2 V^2))$. Here $W(T)$ and $Q(T)$ are free parameters for each temperature and gate voltage point. The average values of the coefficients across the Kondo plateau were α = 0.12 ± 0.015 and γ = 0.7 ± 0.1 for both fits. Though out of the functions we considered the modified quadratic fits the experimental data over the widest bias range (±15 μV in the center of the Kondo plateau), there is little physical basis for using this fit. Likewise, even though a Lorentzian is commonly used for fitting Kondo peaks in the experimental literature, existing theoretical work indicates that the true peak shape is more complex.

### 4.2 Effects of Allowing T-independent Conductance Offset in $G_0$ and $T_K$ Fits

Even though we believe $G_0$ and $T_K$ values extracted by including a temperature independent conductance offset do not accurately reflect behavior in the low temperature limit, it is worth noting how the coefficient values are modified by the use of an offset, due to the prevalence of its usage in the experimental literature. The scaling factors $G_0$ and $T_K$ are both smaller when an offset is included, resulting in an overall decrease in the extracted values of α and γ. The coefficients retain the same qualitative behavior across the Kondo plateau as seen in Fig. 3 of the main text, but the average values of α = 0.08± 0.015 and γ = 0.24± 0.07. The scaling exponents are not affected by the use of an offset.